\documentclass[aps,prl,reprint,superscriptaddress,nobibnotes]{revtex4-2}
\usepackage{graphicx}
\usepackage{array,multirow}
\usepackage{amsmath}
\usepackage{xcolor, soul}
\usepackage{cancel}
\usepackage{comment}
\usepackage{hyperref}
\usepackage{wasysym}
\usepackage[normalem]{ulem}
\usepackage[utf8]{inputenc}
\begin{document}

\title{Improved Limit on Tensor Currents in the Weak Interaction from $^8\text{Li}$ $\beta$ Decay}

\author{M. T. Burkey}

\affiliation{Lawrence Livermore National Laboratory, Livermore, California 94550, USA}
\affiliation{Department of Physics, University of Chicago, Chicago, Illinois 60637, USA}
\affiliation{Physics Division, Argonne National Laboratory, Lemont, Illinois 60439, USA}
\author{G. Savard}
\affiliation{Department of Physics, University of Chicago, Chicago, Illinois 60637, USA}
\affiliation{Physics Division, Argonne National Laboratory, Lemont, Illinois 60439, USA}
\author{A. T. Gallant}
\affiliation{Lawrence Livermore National Laboratory, Livermore, California 94550, USA}
\author{N. D. Scielzo}
\affiliation{Lawrence Livermore National Laboratory, Livermore, California 94550, USA}
\author{J. A. Clark}
\affiliation{Physics Division, Argonne National Laboratory, Lemont, Illinois 60439, USA}
\affiliation{Department of Physics and Astronomy, University of Manitoba, Winnipeg, Manitoba R3T 2N2, Canada}
\author{T. Y. Hirsh}
\affiliation{Department of Physics and Astronomy, University of Manitoba, Winnipeg, Manitoba R3T 2N2, Canada}
\affiliation{Physics Division, Argonne National Laboratory, Lemont, Illinois 60439, USA}
\affiliation{Soreq NRC, Yavne 81800, Israel}
\author{L. Varriano }
\affiliation{Department of Physics, University of Chicago, Chicago, Illinois 60637, USA}
\affiliation{Physics Division, Argonne National Laboratory, Lemont, Illinois 60439, USA}
\author{G. H. Sargsyan}
\affiliation{Department of Physics and Astronomy, Louisiana State University, Louisiana 70803, USA}
\author{K. D. Launey}
\affiliation{Department of Physics and Astronomy, Louisiana State University, Louisiana 70803, USA}
\author{M. Brodeur}
\affiliation{Department of Physics, University of Notre Dame, Notre Dame, Indiana 46556, USA}
\author{D. P. Burdette}
\affiliation{Department of Physics, University of Notre Dame, Notre Dame, Indiana 46556, USA}
\affiliation{Physics Division, Argonne National Laboratory, Lemont, Illinois 60439, USA}
\author{E. Heckmaier}
\affiliation{Department of Physics and Astronomy, University of California Irvine, Irvine, California 92697, USA}
\affiliation{Lawrence Livermore National Laboratory, Livermore, California 94550, USA}
\author{K. Joerres}
\altaffiliation{Present Address: Wolfram Research, Champaign, IL 61820, USA}
\affiliation{Department of Physics and Astronomy, Louisiana State University, Louisiana 70803, USA}
\author{J. W. Klimes}
\altaffiliation{Present Address: GSI Helmholtz Center for Heavy Ion Research, 64291 Darmstadt, Germany}
\affiliation{Physics Division, Argonne National Laboratory, Lemont, Illinois 60439, USA}
\author{K. Kolos}
\affiliation{Lawrence Livermore National Laboratory, Livermore, California 94550, USA}
\author{A. Laminack}
\altaffiliation{Present Address: Physics Division, Oak Ridge National Laboratory, Oak Ridge, TN 37831, USA}
\affiliation{Department of Physics and Astronomy, Louisiana State University, Louisiana 70803, USA}
\author{K. G. Leach}
\affiliation{Department of Physics, Colorado School of Mines, Golden, Colorado, 80401 USA}
\author{A. F. Levand}
\affiliation{Physics Division, Argonne National Laboratory, Lemont, Illinois 60439, USA}
\author{B. Longfellow}
\affiliation{Lawrence Livermore National Laboratory, Livermore, California 94550, USA}
\author{B. Maa{\ss}}
\affiliation{Institut f\"{u}r Kernphysik, Technische Universit\"{a}t Darmstadt, 64289 Darmstadt, Germany}
\affiliation{Physics Division, Argonne National Laboratory, Lemont, Illinois 60439, USA}
\author{S. T. Marley}
\affiliation{Department of Physics and Astronomy, Louisiana State University, Louisiana 70803, USA}
\author{G. E. Morgan}
\affiliation{Department of Physics and Astronomy, Louisiana State University, Louisiana 70803, USA}
\author{P. Mueller}
\affiliation{Physics Division, Argonne National Laboratory, Lemont, Illinois 60439, USA}
\author{R. Orford}
\altaffiliation{Present Address: Nuclear Science Division, Lawrence Berkeley National Laboratory, Berkeley, CA 94720}
\affiliation{Physics Division, Argonne National Laboratory, Lemont, Illinois 60439, USA}
\affiliation{Department of Physics, McGill University, Montr\'{e}al, Qu\'{e}bec H3A 2T8, Canada }
\author{S. W. Padgett}
\altaffiliation{Present Address: Peraton  Inc., Colorado Springs, CO 80919}
\affiliation{Lawrence Livermore National Laboratory, Livermore, California 94550, USA}
\author{A. P\'{e}rez Galv\'{a}n}
\altaffiliation{Present Address: Vertex Pharmaceuticals, San Diego, CA 92121, USA}
\affiliation{Physics Division, Argonne National Laboratory, Lemont, Illinois 60439, USA}
\author{J. R. Pierce}
\altaffiliation{Present Address: Department of Physics and Astronomy, University of California, Los Angeles, CA 90095, USA}
\affiliation{Department of Physics, University of Chicago, Chicago, Illinois 60637, USA}
\affiliation{Physics Division, Argonne National Laboratory, Lemont, Illinois 60439, USA}
\author{D. Ray}
\affiliation{Department of Physics and Astronomy, University of Manitoba, Winnipeg, Manitoba R3T 2N2, Canada}
\affiliation{Physics Division, Argonne National Laboratory, Lemont, Illinois 60439, USA}
\author{R. Segel}
\affiliation{Department of Physics and Astronomy, Northwestern University, Evanston, Illinois 60208, USA}
\author{K. Siegl}
\altaffiliation{Present Address: Department of Physics and Astronomy, University of Tennessee, Knoxville, TN 37996}
\affiliation{Department of Physics, University of Notre Dame, Notre Dame, Indiana 46556, USA}
\author{K. S. Sharma}
\affiliation{Department of Physics and Astronomy, University of Manitoba, Winnipeg, Manitoba R3T 2N2, Canada}
\author{B. S. Wang}
\affiliation{Lawrence Livermore National Laboratory, Livermore, California 94550, USA}

\date{\today , LLNL-JRNL-823604}

\begin{abstract}
The electroweak interaction in the Standard Model (SM) is described by a pure vector-axial-vector structure, though any Lorentz-invariant component could contribute. In this work, we present the most precise measurement of tensor currents in the low-energy regime by examining the $\beta$-$\bar{\nu}$ correlation of trapped $^{8}$Li ions with the Beta-decay Paul Trap. We find $a_{\beta\nu} = -0.3325 \pm 0.0013_{stat} \pm 0.0019_{syst}$ at $1\sigma$ for the case of coupling to right-handed neutrinos $(C_T=-C_T')$, which is consistent with the SM prediction.
\end{abstract}

\maketitle
Measurements of angular correlations in nuclear $\beta$ decay are well suited and widely used  to test the electroweak interaction Standard Model (SM) description while also serving as a broadband test for new physics \cite{GONZALEZ2019, RevModPhys.78.991}. Though data presently favors only vector $(V)$ and axial-vector $(A)$ couplings in the electroweak Lagrangian, the other Lorentz-invariant interactions [Scalar $(S)$, Tensor $(T)$, and Pseudoscalar $(P)$] can arise in SM extensions, such as leptoquark exchanges and contact interactions \cite{HERCZEG2001413}. The coupling constants are defined as $C_i$ for ``parity-even" interactions and $C_i'$ for ``parity-odd" interactions ($i=\text{S}$, $V$, $T$, $A$, or $P$), with parity maximally violated $(C_i=C_i')$ in the SM. The $\beta\text{-}\bar{\nu}$ correlation coefficient $a_{\beta\nu}$ correlates the directions of the emitted leptons in $\beta$ decay and is dependent on the coupling constants. For pure Gamow-Teller $(A)$ decays, and pure Fermi $(V)$ decays, $a_{\beta\nu}$ is expected to be $-1/3$ and $+1$, respectively. Non-SM interactions would lead to deviations from these values.

The development of intense, low-energy beams of radioactive nuclei has greatly aided the current generation of $\beta\text{-}\bar{\nu}$ angular correlation experiments with ion and atom traps \cite{Argon35, 21Na, LPC-GANIL, CENPA, Mukul:2017myi}. Traps are an ideal tool for these measurements, as the nuclide of interest is held nearly at rest in a small, well-characterized volume at high vacuum. This allows the decay products to propagate to an array of detectors with minimal scattering. Thus, complete $\beta\text{-decay}$ kinematic reconstruction can be achieved, which enables $\beta\text{-}\bar{\nu}$ correlation measurements to pursue $\text{sub-}1\%$ precision. 

The highest precision nuclear $\beta\text{-}\bar{\nu}$ correlation limits on $T$ currents were set from a corrected $^6\text{He}$ $\beta\text{-decay}$ measurement from 1963: $a_{\beta\nu}=-0.3308 \pm 0.0030$ \cite{He6, gluckhe6} and our previous $^{8}\text{Li}$ work: $a_{\beta\nu}=-0.3342 \pm 0.0038$ \cite{Matt8Li}, both of which involve Gamow-Teller decays. In 2019, a global analysis of available neutron and nuclear $\beta$-decay data estimated $0.003{<}|C_T/C_A|{<}0.078$ $(|C_T/C_A|^2 \lesssim 0.0061)$ \cite{GONZALEZ2019} at $95.5\%$ CL, with the assumption of right-handed couplings for tensor currents $(C_T=-C_T'$ and $b_{Fierz}=0)$. Here, for the purpose of discussion we use the same simplification. When lifted, the $a_{\beta\nu}$ result becomes $\tilde{a}_{\beta\nu}=a_{\beta\nu}/(1+b_{Fierz}\langle m_e / E \rangle)$, where $E$ is the $\beta$ energy. The global analysis was updated in 2021 by Falkowski \textit{et al.} \cite{Falkowski2021} to include a 2020 aSPECT neutron decay measurement, which pushed the total right-handed tensor current strength from $+1.8\sigma$ to $+3.2\sigma$ away from the SM. \cite{aspect}. High-energy measurements at the Large Hadron Collider provide tensor-current limits that are comparable \cite{Falkowski2021} or in the case of right-handed couplings, more stringent \cite{GONZALEZ2019} than those achieved from $\beta$ decay, although substantially different energy scales and assumptions are required.

This Letter presents an improved limit on $T$ contributions obtained from a high-precision study of $^{8}\text{Li}$ $\beta$ decay performed with the Beta-decay Paul Trap (BPT) \cite{BPTpaper}. The experimental setup and data analysis are built upon our earlier efforts to study $^{8}\text{Li}$ \cite{GangLi, Matt8Li}.

\begin{figure}
\includegraphics[width=\linewidth]{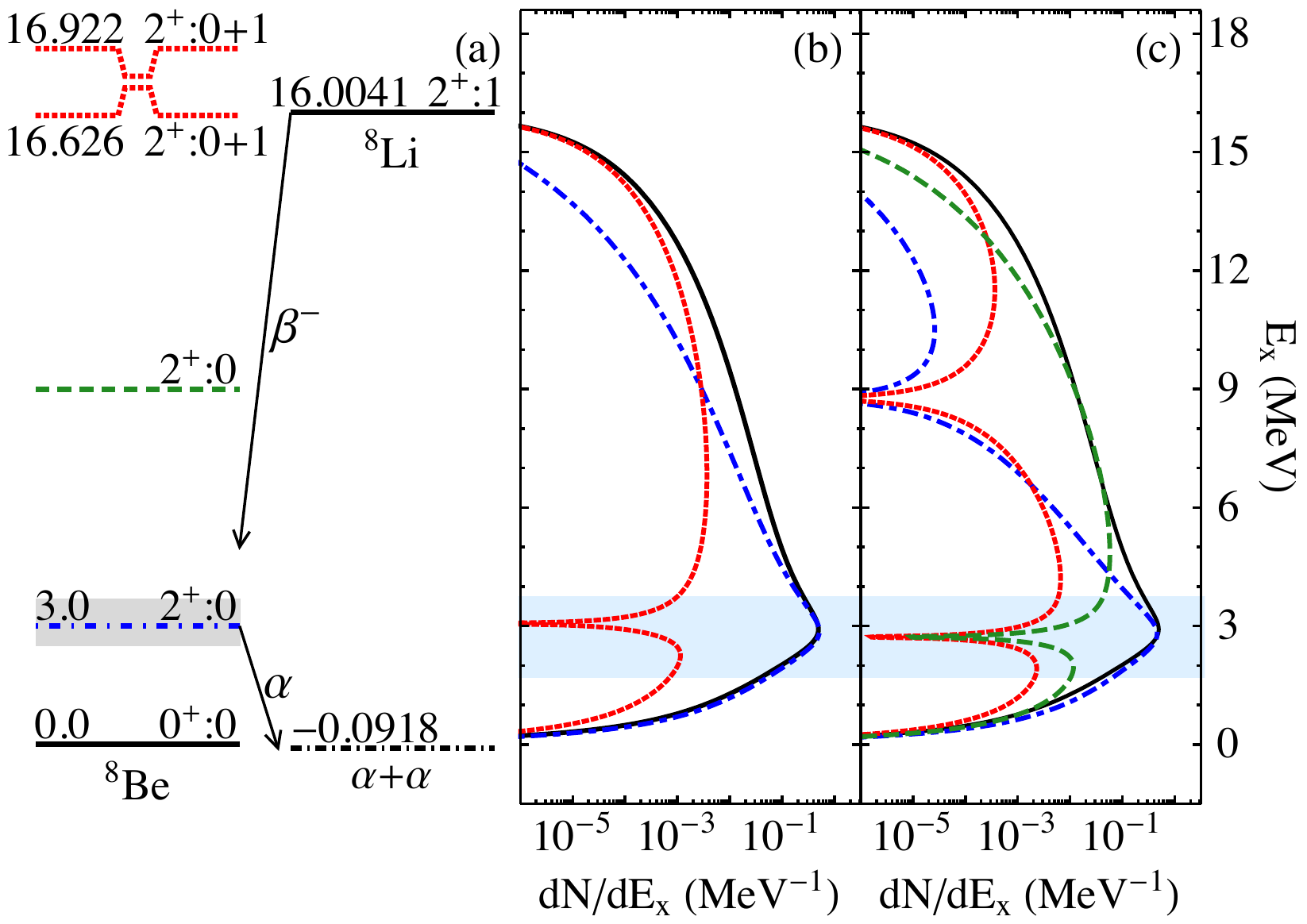}
\caption{ (a) The $\beta\text{-decay}$ scheme of $^8\text{Li}$ (with $E_x$, $J^{\pi}$, and $\text{T}$ listed from left to right) \cite{TILLEY2004}. Sample $^{8}\text{Be}$ excitation energy $E_x$ spectrum R-matrix fits (black) with approximate individual state contributions from (b) the 3-MeV resonance (blue) and doublet states (red) and (c) with an intruder state (green) centered at ${\sim} 9\text{ MeV}$ added. The region highlighted in blue is the $E_x$ range used in our analysis.  \label{fig:rmatrix}}
 \end{figure}

The decay of $^{8}\text{Li}$ is ideal for $\beta\text{-decay}$ angular correlation measurements in an ion trap, due to its nearly-pure Gamow-Teller transition from the $J^{\pi}=2^{+}$, isospin $\text{T}=1$ ground state to a broad $J^{\pi}=2^{+}$, $\text{T}=0$ $^{8}\text{Be}$ excited state that immediately breaks apart into two $\alpha$ particles (see Fig. \ref{fig:rmatrix}[a]). \textit{Ab initio} calculations indicate that the Fermi contribution to the 3-MeV-resonance matrix element is ${<}10^{-3}$ \cite{PhysRevC.88.044333} and the nearest Fermi-decay strength is centered closely around the doublet transition between 16 and 17 MeV (``doublet," hereafter). Both contributions are below our experimental sensitivity. 

In the allowed approximation, the $^{8}\text{Li}$ decay rate can be expressed as \cite{Holstein}:

\begin{align}
d\Gamma &\propto F(Z, E)p_e E(E_0-E)^2\bigg[g_1+g_2\frac{\vec{p}_e}{E} \cdot \hat{p}_{\bar{\nu}}  \nonumber \\ &+ g_{12}
\bigg(\Big[\hat{p}_{\alpha} \cdot \frac{\vec{p}_e}{E}\Big] \big[\hat{p}_{\alpha} \cdot \hat{p}_{\bar{\nu}}\big] -\frac{1}{3}\frac{\vec{p}_e}{E}\cdot \hat{p}_{\bar{\nu}} \bigg) \bigg]
\end{align}

\noindent where $E_0$ and $(E$, $\vec{p}_e)$ are the $\beta$ endpoint energy and four-momentum, $\hat{p}_{\alpha}$ and $\hat{p}_{\bar{\nu}}$ are the $\alpha$ and $\bar{\nu}$ momentum unit vectors, respectively, and $F(Z,E)$ is the Fermi function. The $g_i$ terms are spectral functions dependent on the $C_i\text{'s}$, and to a lesser degree, $E$, $E_0$, and several recoil-order form factors: the weak magnetism term $b$, the induced tensor term $d$ \footnote{The induced tensor term $d$ arises as a higher-order recoil term and is separate from the intrinsic tensor currents $C_T$ that are the subject of this study} , and the second-forbidden axial-vector terms $j_2$, $j_3$. These recoil-order corrections also give rise to additional correlations between the $\beta$, $\bar{\nu}$, and $\alpha$ particles that are ${\sim} 100 \times$ smaller than the terms shown in Eq. 1.

The triple-correlation term that arises from the delayed $\alpha$ emission can be exploited to increase sensitivity to $a_{\beta\nu} \equiv g_2/g_1$. When the $\beta$ and an $\alpha$ particle are emitted in the same direction, the angular correlation factor of $g_{12}$ becomes $\frac{2}{3}(\hat{p}_e \cdot \hat{p}_{\bar{\nu}})$, resulting in $a_{\beta\nu}^{\text{eff}} = (g_2 + \frac{2}{3}g_{12})/g_1$. In the $^{8}\text{Li}$ decay spin sequence, $g_{1}{=}1, \: g_{2}{=}-1/3$, and $g_{12}{=}-1$. Thus, by selecting approximately parallel $\alpha\text{-}\beta$ events, the measurement's sensitivity to the $\beta\text{-}\bar{\nu}$ angular correlation increases by up to $3\times$. Further, due to the large $Q_{\beta}$ (16.00413(6) MeV \cite{Wang_2017}) and small nuclear mass, the $^8\text{Be}^*$ recoil energy is comparatively large, resulting in kinematic shifts that produce ${\sim}400\text{ keV}$ $\alpha$-particle energy differences $\Delta E_{\alpha}$ in the lab frame. $\Delta E_{\alpha}$ is straightforward to measure and is influenced by $a_{\beta\nu}$.

The decay of $^{8}\text{Li}$ populates a broad excitation energy spectrum, which leads to some complications. In Fig. \ref{fig:rmatrix}, the $^8\text{Li}$ level scheme is shown alongside R-matrix fits of the $^{8}\text{Be}$ excitation energy $E_x$ spectrum obtained from this data (similar to the fits in Refs. \cite{Barker2, Warburton}) with approximate individual state contributions to the spectrum \cite{8He}. Though the doublet states are above $Q_{\beta}$, their Gamow-Teller matrix elements are large and their resonance tails extend to significantly lower energies. The decay strength to the doublet increases with $E_x$, eventually dominating the transitions at $E_x{>}10$ MeV. Furthermore, the 3-MeV and doublet transitions each have significantly different recoil-order form factors that must be considered. While the state-dependent recoil-order contributions are interesting in their own right \cite{McKeown,Tribble}, here we minimize these effects by focusing on transitions to $E_x {\sim} 3\text{ MeV}$ (the shaded area in Fig. \ref{fig:rmatrix}[b-c]). 

In our previous work \cite{Matt8Li}, the recoil-order form factors were taken from results in Sumikama \textit{et al.} \cite{j2j3}. Due to statistical constraints and the recoil effects' small size, the measured form factors  obtained from that work were averaged over the entire $^8\text{Be}$ $E_x$ spectrum and had comparatively large uncertainties. Utilizing \textit{ab initio} symmetry-adapted no-core shell model (SA-NCSM) calculations \cite{DytrychLDRWRBB20, LauneyDD16} correlated to the measured $^8\text{Li}$ ground-state quadrupole moment, more precise values of the form factors for each relevant $^{8}\text{Be}$ transition have been determined  \cite{Grigor} and were used here. With the exception of $b$, the values from Ref. \cite{j2j3} were approximately halfway between the 3-MeV and doublet transitions' calculated form factors and all associated uncertainties of the 3-MeV transition values were constrained to within $10\%.$

In addition to reproducing the known $^{8}\text{Be}$ states, the SA-NCSM calculations also predict a low-lying, $2^{+}$ $\alpha+\alpha$ state with a width of $10(3)$ MeV (calculated using a $\text{NNLO}_{\text{opt}}$ chiral potential) \cite{Grigor}, which would be accessible to $^8\text{Li}$ via allowed $\beta\text{ decay}$. There has been an ongoing debate about the existence of this so-called ``intruder state," though experimental evidence remains inconclusive \cite{Warburton, TILLEY2004, Barker1, BARKER1994, HUMBLET1998,Fayache,MUNCH2018779, FayacheErratum, Barker2000, Caurier}. This measurement was also unable to reach a conclusion on the intruder resonance's existence based on R-matrix fitting. An R-matrix fit including a $2^{+}$ intruder state is shown in Fig. \ref{fig:rmatrix}(c). Due to the interference between the lowest two broad states, the intruder state would contribute to the transition strength between ${\sim}3\text{-}15\text{ MeV}$, which introduces some minor systematic uncertainty in the $E_x$ range used in our angular-correlation analysis. More details on the intruder-state systematic will be discussed with the other uncertainties and our R-matrix fitting will be covered in a future publication. 

A description of the experimental apparatus can be found in Refs. \cite{Matt8Li, GangLi}. Only key details and changes since the previous experiment \cite{Matt8Li} will be covered here. The ion production and transport at the Argonne Tandem-Linac Accelerator System (ATLAS) was modified to more efficiently produce $^7\text{Li}^+$, and the beam-line used to transport the ions after the $^7\text{Li}(d, p)^8\text{Li}$ reaction was outfitted with a new gas catcher \cite{Guygascatcher} and beamstop. These changes resulted in an order-of-magnitude increase in the rate of $^8\text{Li}^+$ ions delivered to the BPT compared to our previous experiment \cite{Matt8Li}. 

\begin{figure}
\includegraphics[width=\linewidth]{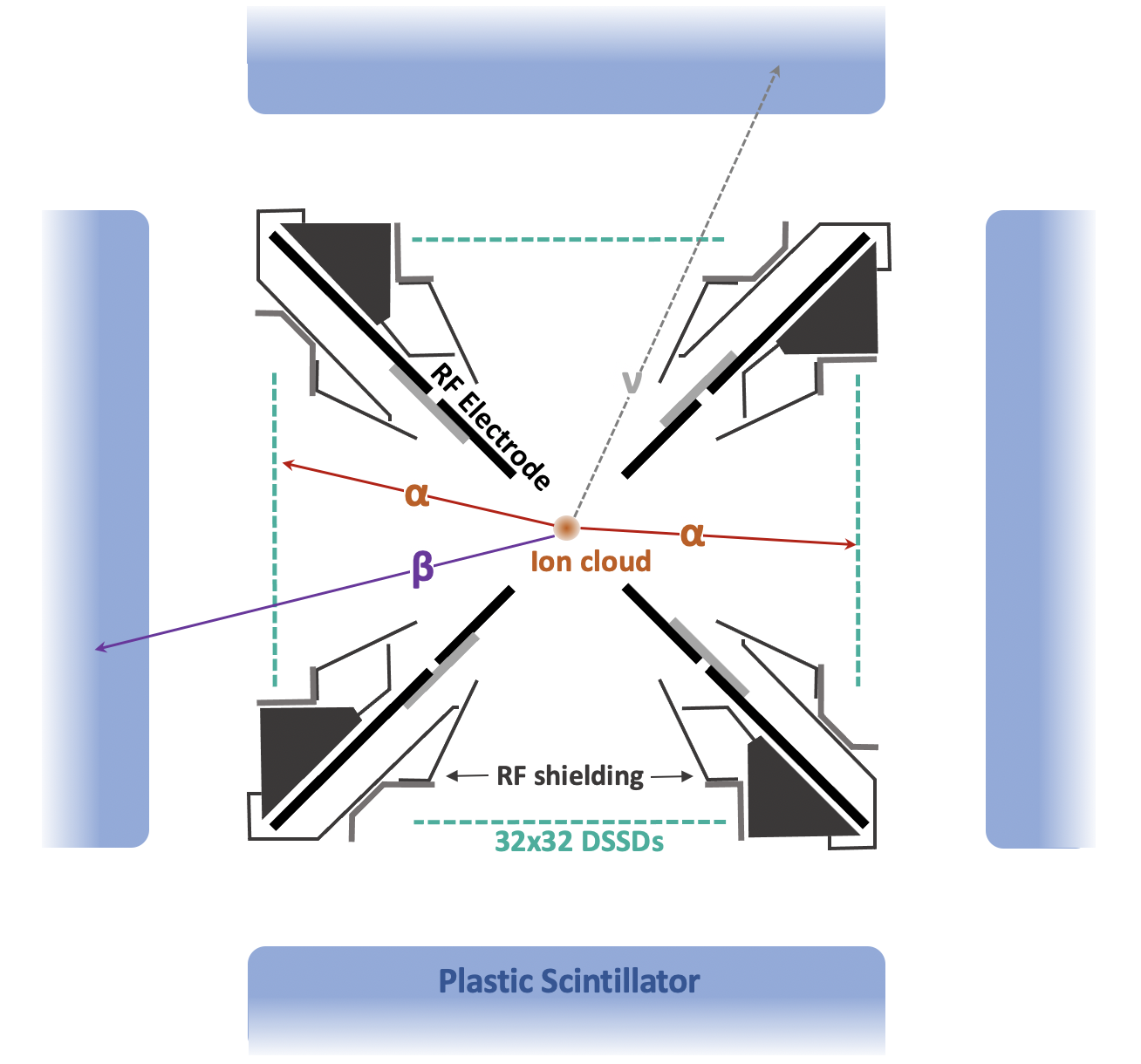}
\caption{Radial-plane cross-sectional view of the BPT showing a typical triple event. \label{fig:bpt}}
 \end{figure}

The BPT, shown schematically in Fig. \ref{fig:bpt} is a linear Paul trap with thin, segmented, planar electrodes that confine $^{8}\text{Li}^+$ ions within a small $({\sim}1\text{ mm}^3)$ volume at the trap center. The BPT utilizes a combination of radio-frequency (RF) voltage (400 $\text{V}_{pp}$ at 1.3 MHz) and a static DC quadratic potential well with coefficient ${\sim} 3\text{ V/cm}^2$ to provide radial and axial confinement, respectively. The ions are cooled through interaction with a high-purity helium buffer gas at a pressure of $10^{-5}$ Torr. The trap frame is cooled to $100\text{ K}$ via liquid nitrogen to improve ion confinement and reduce leakage current in the detectors. Four $64{\times}64{\times}1\text{ mm}^3$ double-sided silicon strip detectors (DSSDs) \cite{micron}, each with 32 strips on the front and back sides, surround the trap. From the struck pixels, both $\alpha$ energies $(E_{\alpha 1}$ and $E_{\alpha 2})$, $\hat{p}_{\alpha 1}$, $\hat{p}_{\alpha 2}$, and $\hat{p}_e$ can be determined. The $\beta\text{-}\alpha\text{-}\alpha$ coincidence signature effectively eliminates all background events. The DSSDs are also bordered by stainless-steel shielding to minimize pickup from the RF voltage applied to the nearby trap electrodes and backed by plastic scintillator detectors \cite{eljen} $(6\text{"}\times 6.2\text{"}\diameter )$ to collect the remaining $\beta$ energy.

Several upgrades to the BPT have been implemented since the experiment in Ref. \cite{Matt8Li}. Tunable notch filters for every DSSD front strip were added before the preamplifiers to remove remaining RF pickup. Of the 128 front strips, only signals from the eight edge strips and an additional five strips were consistently unusable. The \textit{in situ} $^{148}\text{Gd}$ and $^{244}\text{Cm}$ calibration sources were upgraded to a set of spectroscopy-grade sources, which provide $\alpha\text{-particle}$ lines at 3182.690(24) keV \cite{AKOVALI19981} and 5804.77(5) keV \cite{SINGH20082439}, with 20-keV full width at half-maximum \cite{ezag}. 

Over the 14-day experiment, an average of ${\sim}100$ trapped $^{8}\text{Li}$ ions were maintained in the BPT. Events were designated a ``double" when two particles within the same 15-$\mu\text{s}$ event window were detected on opposing DSSDs with deposited energies between 700 and 8000 keV (an $\alpha\text{-}\alpha$ coincidence), while ``triple" events required an additional $\beta$ particle concidence with deposited energy between 200 and 700 keV. The 700-keV threshold was chosen based on \textsc{Geant4} \cite{geant4} simulations of the $\alpha$ and $\beta$ spectra compared to data. 

The DSSD $\alpha$-energy response was calibrated following the method developed in Ref. \cite{Tsviki}, utilizing the $^{148}\text{Gd}$ and $^{244}\text{Cm}$ $\alpha$ lines alongside the DSSD minimum ionizing $\beta$ spectra from the $^{8}\text{Li}$ decay, which served as a low-energy point. The $\beta$ minimum ionizing spectra was matched to \textsc{Geant4} simulations and cross-checked for consistency with cosmic muon data. Following Ref. \cite{lennard}, the calibrated energies were corrected for the detector dead layer, nonionizing energy losses (NIELs), and the silicon energy-response non-linearity \cite{lennard, phd-k}. The data-collection system non-linearity was also accounted for \cite{HPInt}.

After calibration, several cuts were applied. (i) Coincidences detected less than 30 ms after a new ion bunch is injected into the cloud were discarded, as opening the trap briefly disturbs the ion cloud's thermal equilibrium. (ii) Both $E_{\alpha 1}$ and $E_{\alpha 2}$ must be greater than 850 keV to accommodate the aforementioned calibration corrections. (iii) $E_{\alpha 1}+E_{\alpha 2} < 3.75\text{ MeV}$ (note: $E_x= E_{\alpha 1}+E_{\alpha 2}+91.2\text{ keV}-E_{\text{recoil}}^{\text{Be}})$ to minimize uncertainty associated with the possible existence of an intruder resonance. (iv) The difference in recorded $\alpha$ energy between the front and back strips must be within 30 keV, which eliminates most $\alpha$ particle events that interact with the inter-strip gap between front strips, where charge is not fully collected. 

This analysis focused on triple coincidences where the $\beta$ hit one of the detectors struck by an $\alpha$ particle, allowing for the increase in sensitivity to $a_{\beta\nu}$. Taking into account all these constraints, the final number of triples used for analysis was $2.9{\times}10^5$, amounting to ${\sim}1\%$ of all $^8\text{Li}$ decays in the BPT.

Our data were compared to a detailed simulation of the decay kinematics and experimental system \cite{Matt8Li, HPInt}. The decay is generated via Monte-Carlo sampling of the $\beta$-delayed $\alpha$ emission phase space \cite{Holstein, nickBD, radius, WILKINSON1990509,WILKINSON1993182}. The $^{8}\text{Be}^{*}$ final-state distribution is obtained from an R-matrix fit to the calibrated $^{8}\text{Li}$ data. Radiative corrections based on Gl\"{u}ck's methodology are included \cite{GLUCK1997}. The $\beta$ particles' deposited energies are determined with a detailed \textsc{Geant4} simulation using the ``option3" standard electromagnetic physics list \cite{geant4,urban,lewistheory}. The geometry of the trap and detector array were imported into GEANT using a GDML-adapted \cite{instep} BPT design developed in Autodesk Inventor \cite{Autodesk}. 

The simulation propagates the $\alpha$ particles to their projected detector hit locations, and the simulated $E_{\alpha}$ values are passed through an algorithm that applies a randomized shift to account for the energy-dependent DSSD response or ``lineshape." The lineshape distribution was constructed using calibration-source and beam measurements alongside the detector manufacturer's specifications for the inactive dead layer and the charge-collecting aluminum strips mounted on deep silicon implants framing each strip. Refs. \cite{Tsviki, lineshape} contain the lineshape convolution methodology.

\begin{figure}
\includegraphics[width=\linewidth]{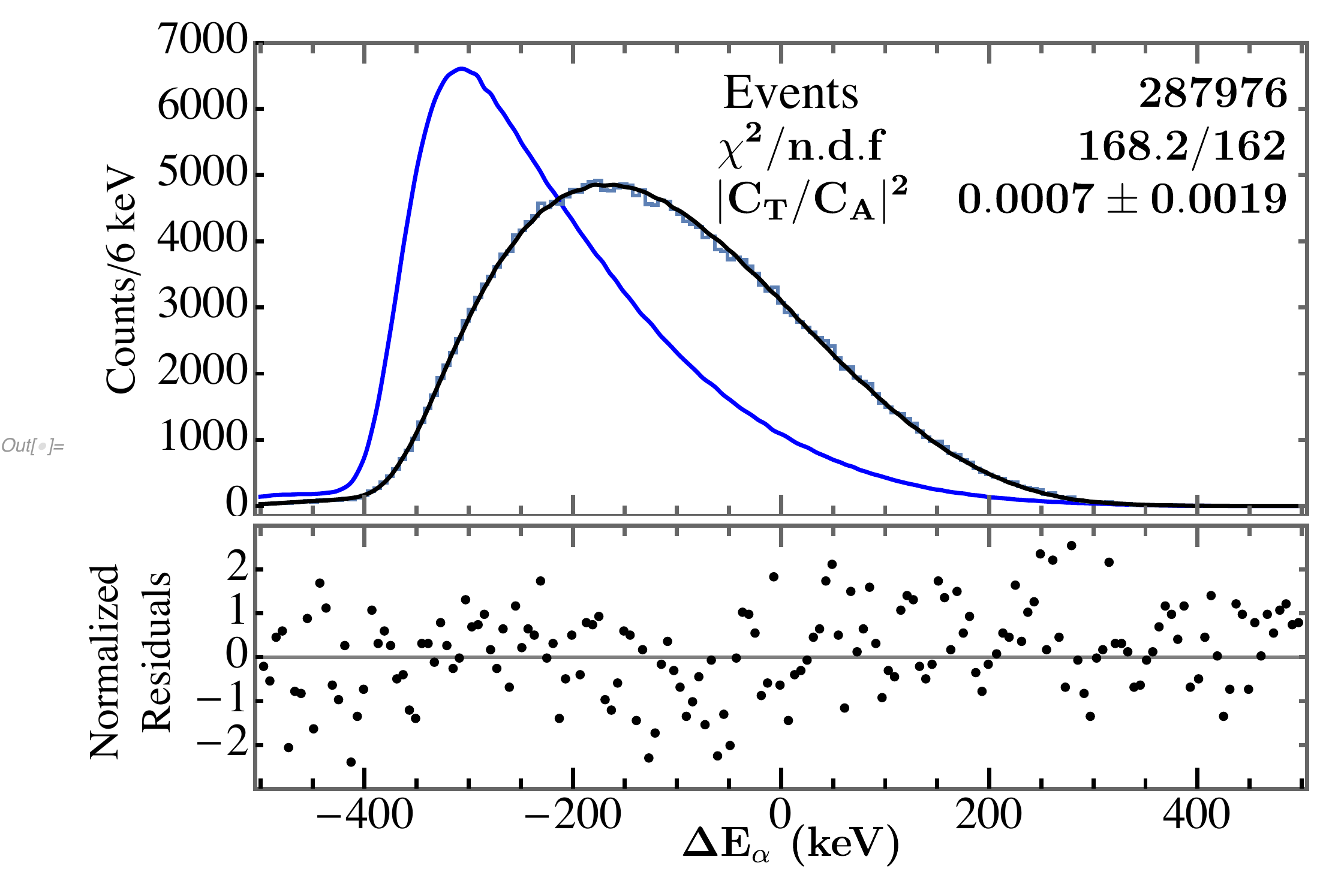}
\caption{Measured $\Delta E_{\alpha}$ spectrum fit to a linear combination of simulated pure $A$ and $T$ interactions (black curve) with the normalized residuals below. For comparison, an example pure $T$ interaction spectrum is plotted in blue. \label{fig:alphadiff}}
 \end{figure}

For events where the $\beta$ and $\alpha$ strike the same detector, $T$ interactions result in larger average recoil energies than $A$ interactions due to the alignment of the lepton momenta. The recoil energy is observed through the kinematic shifts resulting in $\Delta E_{\alpha}$; consequently we are able to sensitively extract $|C_T/C_A|^2$ from the $\Delta E_{\alpha}$ spectrum. Spectra for a pure $A$ and a pure $T$ interaction are generated with our simulation. The data are then fit to a linear combination of the two spectra, with the relative amplitudes of couplings, $|C_T/C_A|^2$, and the normalization as the only fitting parameters \cite{Matt8Li}. The experimental results and the best fit to the data are shown in Fig. \ref{fig:alphadiff}. 

Table \ref{tab:totalsys} summarizes the dominant systematic uncertainties at $1\sigma$ for $|C_T/C_A|^2$. The total is calculated by summing the components in quadrature, with the exception of the intruder state, which is added in linearly at the end. The entries of Table \ref{tab:totalsys} are briefly explained below. 

\textit{Intruder state}\textemdash If the $2^{+}$ intruder resonance is present, we estimate from our R-matrix fits that ${\sim}6\%$ of events decay via that transition below the $E_{\alpha 1}+E_{\alpha 2} {<} 3.75\text{ MeV}$ cutoff. Due to differences in the recoil-order terms, the intruder events would increase $|C_T/C_A|^2$ by $+0.0010$. To account for this, we shift our measured $|C_T/C_A|^2$ by half of the intruder state increase and take an uncertainty of $\Delta|C_T/C_A|^2=0.0005$, which spans either case.

\textit{Recoil $\&$ Radiative Terms}\textemdash The uncertainties associated with all the SA-NCSM-calculated form factors in Ref. \cite{Grigor} yielded a total uncertainty on $|C_T/C_A|^2$ of 0.0013, with $d$ being the dominant contributor. This represents a ${>}60\%$ improvement from the systematic uncertainty obtained by using the Sumikama \textit{et al.} results \cite{j2j3}. The uncertainty associated with Z-independent radiative corrections was 0.0008. Summed in quadrature, the two yield a combined uncertainty $\Delta|C_T/C_A|^2=0.0015$.

\textit{$\alpha$-energy calibration}\textemdash The largest contributions arise from several energy corrections during the calibration process: energy lost through the 100-nm-thick DSSD dead layer, fitted distributions of the NIEL generated in TRIM \cite{SRIM}, and the measured silicon energy-response non-linearity parameters (uncertainties taken from Refs. \cite{phd-k, e0}). The combined systematic uncertainty of $|C_T/C_A|^2$ for the $\alpha$-energy calibration is 0.0007.

\textit{Detector lineshape}\textemdash Uncertainties in the lineshape model resulted in a $\Delta|C_T/C_A|^2=0.0009$, of which the largest contribution arose from uncertainty associated with charge sharing across the back strips. 

\textit{Data cuts}\textemdash All of the data cuts were adjusted within reasonable ranges and the resulting uncertainties were added in quadrature; this yielded $\Delta|C_T/C_A|^2=0.0009$, with the dominant contributor being the 700-keV threshold used to discriminate between $\alpha$ and $\beta$ particles. 

\textit{$\beta$ scattering}\textemdash Scattering within the trap increases the number of $\beta$ particles striking the DSSDs and distorts the angular correlation for those extra triple events. Both the triple events/double events ratio (T3/D2) and the backscattered triple events/triple events ratio were consistent between simulation and data, even with much smaller statistical uncertainty, while the plastic detectors assisted with distinguishing between origins of scattering within the trap. The $\beta$-scattering uncertainty was determined by extracting $|C_T/C_A|^2$ using two sets of simulations --- one set with some scattered triple events added and another with some scattered events discarded --- to yield simulated T3/D2 ratios $\pm 2\sigma$ from the measured ratio. The average magnitude of $\Delta|C_T/C_A|^2$ was 0.0010.

Increasing the time reserved for measuring untrapped $^8\text{Li}$ by $7\times$ compared to the 2015 experiment \cite{Matt8Li} reduced the background systematic uncertainty to below our sensitivity. The systematic uncertainties associated with the simulated $^8\text{Be}^{*}$ final-state distribution, and the ion-cloud characteristics were also negligible.

\begin{table}
\caption{Summary of dominant systematic corrections and uncertainties, listed at $1\sigma$.  \label{tab:totalsys}}
\begin{ruledtabular}
\renewcommand{\arraystretch}{1.35}
\begin{tabular}{ l  l c c }
\multicolumn{2}{l}{Source} & Correction & Uncertainty \\ \hline
\parbox[t]{3mm}{\multirow{2}{*}{\rotatebox[origin=c]{90}{{\scriptsize Theory}}}} & Intruder State (added linearly) & +0.0005 & 0.0005 \\
& Recoil $\&$ Radiative Terms & & 0.0015 \\ \hline
\parbox[t]{3mm}{\multirow{4}{*}{\rotatebox[origin=c]{90}{{\scriptsize Experiment}}}}& $\alpha$-Energy Calibration & & 0.0007  \\
&Detector Lineshape &  & 0.0009  \\ 
&Data Cuts & & 0.0009  \\
&$\beta$ Scattering & & 0.0010  \\ \hline
\multicolumn{2}{l}{\textbf{Total}} & +0.0005 & 0.0028 \\
\end{tabular}
\end{ruledtabular}
\end{table}

The result of fitting the $\Delta E_{\alpha}$ spectrum and then
applying the systematic correction was: $|C_T/C_A|^2= 0.0012 \pm 0.0019_{stat} \pm 0.0028_{syst}$ with uncertainties reported at $1\sigma$, which represents a $41\%$ improvement on our previous work's uncertainties and is the single most precise measurement of intrinsic tensor-current contributions to the weak interaction in the low-energy regime. Under the constraint that $C_T=-C'_T$ $(b_{Fierz}\equiv 0)$, $|C_T/C_A|^2$ corresponds to:
$$
a_{\beta\nu} = -0.3325 \pm 0.0013_{stat} \pm 0.0019_{syst}
$$ 
and exceeds the precision of all previous measurements in Gamow-Teller decays. This result can also be interpreted as $|C_T/C_A|^2<0.0076$ or $|C_T/C_A|<0.087$ at the $95.5\%$ CL via a Bayesian analysis with a uniform prior for $|C_T/C_A|^2>0$. If the $C_T=-C'_T$ constraint is lifted, the $1\sigma$ region of possible $|C_T/C_A|$ and $|C_T'/C_A|$ combinations is bounded by the equation: $(|C_T/C_A|+0.044)^2+ (|C_T'/C_A| +0.044)^2 = 0.115^2$, with $\langle m_e / E \rangle=0.0878$. Our findings are in agreement with the SM, in contrast with the global nuclear limits presented in Falwokski \textit{et al.} \cite{Falkowski2021}

Analysis of a similarly-sized data set on the mirror nucleus $^{8}\text{B}$ decay is underway, which will assist with examining the $E_x$-dependency behavior of the decay rate and probing for other non-SM physics, such as deviations from the CVC hypothesis via the weak magnetism term $(b)$. However, an experimental confirmation of the existence of the $2^{+}$ intruder resonance would be highly beneficial to any further investigations in the $A=8$ system. \\

\begin{acknowledgments}
We acknowledge the ATLAS staff for their help and support. This work was carried out under the auspices of the U.S. Department of Energy, by Argonne National Laboratory under Contract No. DE-AC02-06CH11357 and Lawrence Livermore National Laboratory under Contract No. DE-AC52-07NA27344, the National Science Foundation under grants PHY-173857, PHY-2011890, and PHY-1913728, as well as the NSERC, Canada, Application SAPPJ-2018-00028. This research used resources of Argonne National Laboratory's ATLAS facility, which is a DOE Office of Science User Facility. This work also benefited from high performance computational resources provided by LSU,  NERSC (a U.S. DOE Office of Science User Facility operated under Contract No. DE-AC02-05CH11231), as well as the Frontera computing project at TACC (NSF OAC-1818253). M. T. Burkey and L. Varriano were supported by the National Science Foundation Graduate Research Fellowship under Grant Numbers 1144082 and DGE-1746045, respectively. B. Maa{\ss} acknowledges support from the DFG (German Research Foundation) - Project No. 279384907 - SFB 1245.
\end{acknowledgments}

%\bibliography{refs}


\end{document}